\documentclass[a4paper,11pt]{article}
\pdfoutput=1 

\usepackage{jheppub} 
\usepackage[normalem]{ulem}  

\title{Higgs Mechanism in Nonlocal Field Theories}

\author[a]{Manami Noumi Hashi,}
\author[b]{Hiroshi Isono,}
\author[c]{Toshifumi Noumi,}
\author[d]{Gary Shiu,}
\author[e]{and Pablo Soler}

\affiliation[a]{Faculty of Science, Kobe University, Kobe 657-8501, Japan}
\affiliation[b]{Department of Physics, Faculty of Science, Chulalongkorn University, Bangkok 10330, Thailand}
\affiliation[c]{Department of Physics, Kobe University, Kobe 657-8501, Japan}
\affiliation[d]{Department of Physics, University of Wisconsin-Madison, Madison, WI 53706, USA}
\affiliation[e]{Institute for Theoretical Physics, University of Heidelberg, Philosophenweg 19,
D-69120 Heidelberg, Germany}

\preprint{KOBE-COSMO-18-04, MAD-TH-18-02}

\abstract{
We study spontaneous gauge symmetry breaking and the Higgs mechanism in nonlocal field theories. Motivated by the level truncated action of string field theory, we consider a class of nonlocal field theories with an exponential factor of the d'Alembertian attached to the kinetic and mass terms. Modifications of this kind are known to make mild the UV behavior of loop diagrams and thus have been studied not only in the context of string theory but also as an alternative approach to quantum gravity. In this paper we argue that such a nonlocal theory potentially includes a ghost mode near the nonlocal scale in the particle spectrum of the symmetry broken phase. This is in sharp contrast to local field theories and would be an obstruction to making a simple nonlocal model a UV complete theory. We then discuss a possible way out by studying nonlocal theories with extra symmetries such as gauge symmetries in higher spacetime dimensions.
}

\begin{document} 
\setcounter{tocdepth}{2}
\maketitle
\flushbottom

\section{Introduction}
\setcounter{equation}{0}

String theory accommodates a fruitful landscape of low-energy effective field theories, whose understanding is crucial both theoretically and phenomenologically. String field theory~\cite{Witten:1985cc,Zwiebach:1992ie}  provides a field theoretical approach to the structure of the string landscape. In particular, various  non-perturbative aspects of the open string landscape have been clarified both numerically and analytically based on open string field theory (see, e.g, review articles~\cite{Ohmori:2001am,Taylor:2003gn,Fuchs:2008cc,Okawa:2012ica} and references therein).

\medskip
A lesson from the aforementioned developments is that string field theory may describe transitions between vacua with completely different particle contents. A famous example is the phenomenon of open string tachyon condensation: An open string attached to unstable D-branes contains a tachyon in the spectrum. The tachyon vacuum will then describe a vacuum where the unstable D-branes decay and disappear~\cite{Sen:1999xm}. Indeed, open string field theory explicitly showed that there are no particle excitations  in the tachyon vacuum~\cite{Schnabl:2005gv,Ellwood:2006ba}. Such a transition can never be described in ordinary field theories. Then, what makes it possible for string field theory to describe this drastic transition of vacua? First, string field theory contains an infinite number of local fields associated with the infinite types of string excitations. Second, if we write it in terms of local fields, there appear nonlocal terms. For example, the level $0$ sector of open bosonic string field theory~\cite{Witten:1985cc} is given by
\begin{align}
\label{nonlocal_tachyon_intro}
S=\int d^Dx\left[\frac{1}{2}\phi \,e^{-\frac{\Box}{M^2}}(\Box+\mu^2)\phi-\frac{1}{3}g\phi^3\right]
\quad
{\rm with}
\quad
\mu^2>0\,,
\end{align}
where $\phi$ is the tachyon field describing the instability of the D-brane and the exponential factor of the d'Alembertian gives a nonlocality. These two distinctive properties of string field theory are believed to be crucial when one describes the drastic vacuum transition.

\medskip
In this paper we study spontaneous gauge symmetry breaking and the Higgs mechanism in nonlocal field theories, based on the following two motivations: First, in D-brane model building of particle physics, the recombination process of D-branes provides a natural realization of the Standard Model Higgs mechanism~\cite{Cremades:2002cs}. It is a highly stringy process, hence it is desirable if we could follow the whole recombination process in the string field theory framework. Since the analysis in the full string field theory is rather complicated, we would like to take a first step towards such a study by clarifying the nonlocal effect mentioned above in the Higgs mechanism. Second, the exponential factor of the d'Alembertian in Eq.~\eqref{nonlocal_tachyon_intro} is known to make mild the UV behavior of loop diagrams and thus has been studied not only in the context of string theory~\cite{Freund:1987kt,Freund:1987ck,Brekke:1988dg,Ghoshal:2000dd,Minahan:2001pd,Moeller:2002vx,Calcagni:2007wy,Calcagni:2009tx,Calcagni:2009jb,Calcagni:2013eua,Moeller:2003gg,Pius:2016jsl} but also as an alternative approach to quantum field theories and gravity~\cite{Alebastrov:1973np,Moffat:1988zt,Moffat:1990jj,Kato:1990bd, Evens:1990wf,Bakeyev:1996is,Tomboulis:1997gg,Biswas:2005qr,Barnaby:2007ve,Moffat:2010bh,Modesto:2011kw,Biswas:2011ar,Talaganis:2014ida,Biswas:2014yia,Tomboulis:2015gfa,Talaganis:2016ovm,Giaccari:2016kzy,Kimura:2016irk,Ghoshal:2017egr,Calcagni:2018lyd}. It is therefore interesting to study consistency of nonlocal field theories by themselves. Based on these motivations, we discuss a class of nonlocal field theories with an exponential factor of the d'Alembertian. We will find that there appears a drastic change in the particle spectrum of the symmetry broken phase, similarly to the open string tachyon condensation. In particular, we argue that a ghost mode may potentially appear near the nonlocal scale and its absence can be used as a consistency requirement of the theory.

\medskip
This paper is organized as follows. In section~\ref{Sec:tachyon}, we study the tachyon vacuum of the nonlocal model~\eqref{nonlocal_tachyon_intro} to illustrate that the nonlocality affects the particle spectrum of the condensation phase in a significant way. In particular, we show that there appears an undesired ghost mode near the nonlocal scale in a certain range of the parameter space. In Sec.~\ref{Sec:Abelian}, we extend the argument to a nonlocal model with an Abelian gauge field and a complex charged scalar. We find there that the undesired ghost is inevitable in this simple model as long as we consider the parameter region in which the propagators exhibit a mild UV behavior. This observation motivates us to raise a question if nonlocal gauge theories may accommodate a massive gauge boson in a consistent way. In Sec.~\ref{Sec:non-Abelian}, we demonstrate that this is indeed possible if the theory has an extra symmetry. As an illustrative example, we consider a nonlocal Yang-Mills theory with an adjoint Higgs obtained by dimensional reduction of a nonlocal pure Yang-Mills theory in a higher dimensional spacetime. We show that the higher dimensional gauge symmetry constrains the form of nonlocality and consequently the undesired ghost does not show up. We conclude in Sec.~\ref{Sec:conclusion} with a discussion of our results. In Appendix, we provide a concrete form of propagators in the $R_\xi$ type gauge of the model used in Sec.~\ref{Sec:Abelian}.

\subsection*{Note added}

Recently, Ref.~\cite{Gama:2018cda} has appeared on arXiv, which discussed spontaneous symmetry breaking in a nonlocal scalar QED model. The model discussed there is the same as the one we discuss in Sec.~\ref{Sec:Abelian}, but the conclusion is different. First, Ref.~\cite{Gama:2018cda} studied the mass spectrum in the Lorentz gauge rather than the unitary gauge. The mass spectrum obtained this way is thus not physical. Second, the subtlety associated with new additional poles of the propagator near the nonlocal scale was not discussed there. As we discuss in Sec.~\ref{Sec:Abelian}, the additional modes contain a ghost, so that it is a crucial obstruction to making the simple nonlocal model a UV complete theory. One of the purposes of our paper is to clarify this subtlety and explore a possible way out, which is different from the one in Ref.~\cite{Gama:2018cda}.

\section{Tachyon condensation and physical spectra}
\setcounter{equation}{0}
\label{Sec:tachyon}

Let us begin with the following nonlocal action
for a tachyon field $\phi$:\footnote{We use the mostly plus metric $\eta_{\mu\nu}={\rm diag}(-,+,\ldots,+)$ throughout the paper. We also follow the convention of the Feynman rule in Srednicki's textbook~\cite{Srednicki:2007qs}.}
\begin{align}
\label{nonlocal_tachyon}
S=\int d^4x\left[\frac{1}{2}\phi \,e^{-\frac{\Box}{M^2}}(\Box+\mu^2)\phi-\frac{1}{3}g\phi^3\right]
\quad
{\rm with}
\quad
\mu^2>0\,,
\end{align}
where the tachyon mass squared is $m^2_{\rm{tachyon}}=-\mu^2$
and $M$ is the scale characterizing the nonlocality (the local limit is $\mu/M,E/M\to0$ with $E$ being the energy scale of interest). The sign of the exponent is chosen such that the exponential factor induces an exponential suppression of the Euclidean propagator at high energy, which makes mild the UV behavior of loop diagrams. More explicitly, the tachyon propagator is given by
\begin{align}
\Pi_{\phi}=\left[e^{\frac{k^2}{M^2}}(k^2-\mu^2)\right]^{-1}\,,
\end{align}
which has a pole at $k^2=\mu^2=-m^2_{\rm{tachyon}}$. Also, it has an exponential factor $e^{-\frac{k^2}{M^2}}$, which gives a suppression when $k^2\gg M^2$. Such exponential factors arise generically in string field theory~\cite{Ohmori:2001am,Pius:2016jsl}.

\medskip
The homogeneous tachyon vacuum solution
is $\phi_{\rm tv}=\mu^2/g$,
which does not depend on $M$.
The action for the fluctuation $\varphi=\phi-\phi_{\rm tv}$
around the tachyon vacuum is then
\begin{align}
S=\int d^4x\left[\frac{1}{2}\varphi e^{-\frac{\Box}{M^2}}(\Box+\mu^2)\varphi-\mu^2\varphi^2-\frac{1}{3}g\varphi^3\right]\,.
\end{align}
The propagator of $\varphi$ reads
\begin{align}
\label{TV_propagator}
\Pi_\varphi(k)=\left[e^{\frac{k^2}{M^2}}(k^2-\mu^2)+2\mu^2\right]^{-1}\,.
\end{align}
The on-shell condition also follows as
\begin{align}
\label{on-shell}
e^{\frac{k^2}{M^2}}(k^2-\mu^2)+2\mu^2=0\,.
\end{align}
Note that its local limit $M^2\to\infty$ is given by
\begin{align}
(k^2-\mu^2)+2\mu^2=k^2+\mu^2=0\,,
\end{align}
which is simply the on-shell condition with a shifted mass squared.

\medskip
To see how the nonlocality affects the spectrum, let us rewrite Eq.~\eqref{on-shell} as
\begin{align}
f_\lambda(x)=e^{-\lambda x}(x+1)-2=0\,,
\end{align}
where we rescaled $x=-k^2/\mu^2$ and introduced $\lambda=\mu^2/M^2$. In this language, the mass $m$ of the on-shell particle is given by $m^2=x\mu^2$ with $x$ being a solution to $f_\lambda(x)=0$. As depicted in Fig.~\ref{fig:tachyon}, for $\lambda>0$, which is motivated by the sign choice of the exponential factor mentioned above, the function $f_\lambda(x)$ is bounded from above on the real $x$ axis as
\begin{align}
f_\lambda(x)\leq f_\lambda(\lambda^{-1}-1)=\lambda^{-1}e^{\lambda-1}-2\,.
\end{align}
For $\lambda<0$, it is bounded from below instead. Therefore, the number of real solutions to $f_\lambda(x)=0$, i.e., the number of poles of the propagator, depends on the value of $\lambda$. Below let us discuss qualitative properties in the following three branches:
\begin{figure}[t]
\begin{center}
\includegraphics[width=74mm, bb=0 0 360 238]{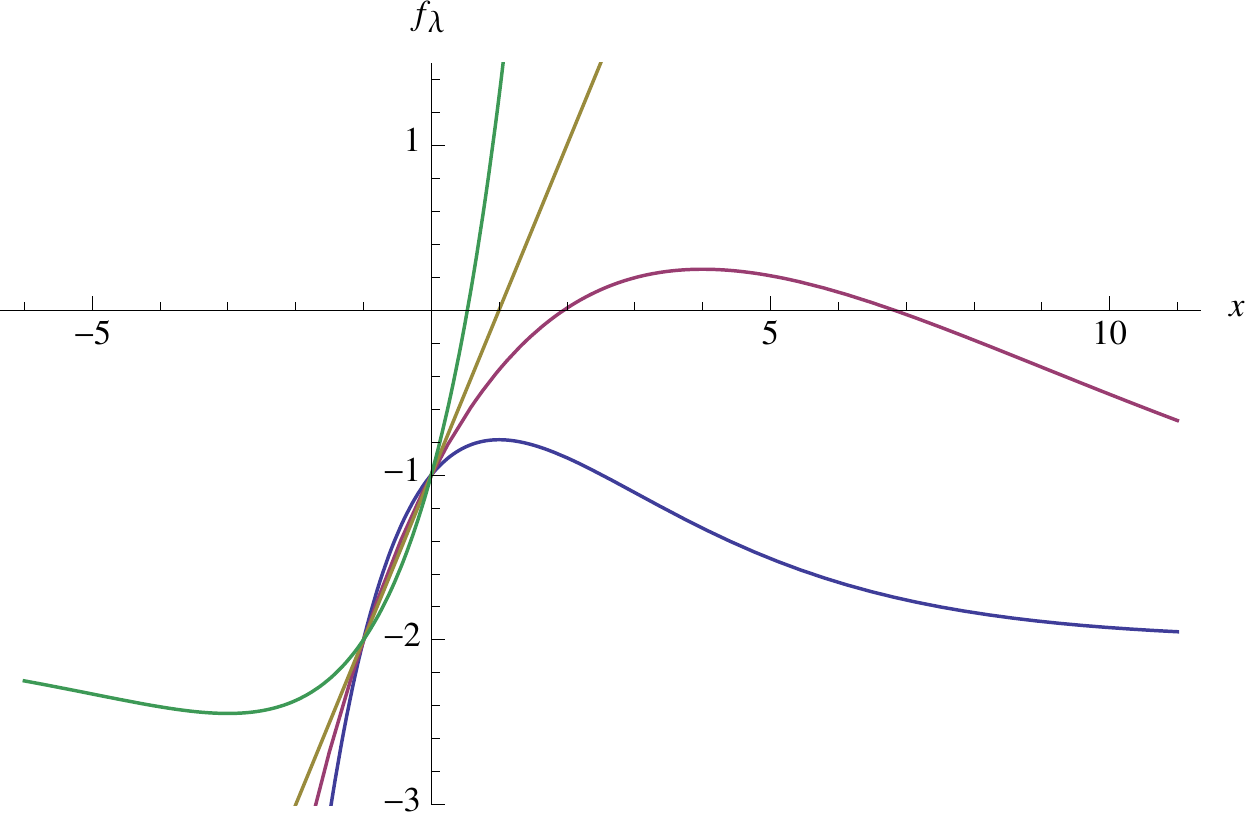}
 \end{center}
\vspace{-5mm}
\caption{
Profile of $f_\lambda(x)$ for $\lambda=0.5$ (blue),
$\lambda=0.2$ (red),
$\lambda=0$ (yellow),
and $\lambda=-0.5$ (green).
In particular, $\lambda=0.5$ has no real solution to $f_\lambda(x)=0$, which implies that there are no physical particle excitations around the tachyon vacuum. Also, $\lambda=0.2$ has two real solutions. One of the two solutions has a negative gradient, which implies that the corresponding mode is a ghost.}
\label{fig:tachyon}
\end{figure}
\begin{enumerate}
\item $\lambda>\lambda_*\simeq0.23$ ($\lambda_*$ is defined through $\lambda_*^{-1}e^{\lambda_*-1}=2$)

In this regime, the function $f_\lambda(x)$ is always negative:
\begin{align}
f_\lambda(x)\leq \lambda^{-1}e^{\lambda-1}-2<0
\,,
\end{align}
so that there are no real solutions to $f_\lambda(x)=0$ and the propagator has no poles on the real $k^2$ axis.\footnote{
Notice however that there exist complex momenta satisfying the on-shell condition, even though their interpretation is not fully understood yet.} We may interpret that there are no physical particle excitations around the tachyon vacuum in this parameter regime. Actually, the same situation appears in the tachyon vacuum of the level $0$ truncated open bosonic string field theory~\cite{Kostelecky:1989nt,Ellwood:2001py}:
In string theory both the nonlocal scale $M$ and the tachyon mass $\mu$ are controlled by the string scale $M_s$, so that the parameter $\lambda=\mu^2/M^2$ is $\mathcal{O}(1)$. More explicitly, the level 0 truncated action of open string field theory is given by\footnote{
In the string field theory context, it is conventional to write the action as
\begin{align}
S=\int d^{26}x\left[\frac{1}{2}\phi(\Box+M_s^2)\phi
-\frac{g_o}{3}\Big(e^{\ln \frac{3\sqrt{3}}{4}\big(\frac{\Box}{M_s^2}+1\big)}\phi\Big)^3\right]\,,
\end{align}
where $g_o$ is the open string coupling. The action~\eqref{OSFT0} is obtained by the redefinition
\begin{align}
\phi\to e^{-\ln \frac{3\sqrt{3}}{4}\big(\frac{\Box}{M_s^2}+1\big)}\phi\,,
\quad
g=\left(\frac{3\sqrt{3}}{4}\right)^3g_o\,.
\end{align}
Note that this field redefinition does not change the on-shell structure because the exponential prefactor does not give rise to any new zero or pole of the d'Alembertian $\Box$.}
\begin{align}
\label{OSFT0}
S=\int d^{26}x\left[\frac{1}{2}\phi\, e^{-\ln\frac{27}{16}\frac{\Box}{M_s^2}}(\Box+M_s^2)\phi
-\frac{g}{3}\phi^3\right]\,,
\end{align}
and hence $\lambda=\ln\frac{27}{16}\simeq0.52>\lambda_*$. In this regime, the propagator has no real poles.
In string theory, this situation is interpreted as the disappearance of open string excitations in the tachyon vacuum due to decay of unstable D-branes~\cite{Sen:1999xm,Schnabl:2005gv,Ellwood:2006ba}.

\item $0<\lambda<\lambda_*$

In this regime, there are two real solutions to $f_\lambda(x)=0$, which correspond to two types of physical particle excitations with the on-shell conditions,
\begin{align}
-k^2=m_1^2
\quad{\rm and}
\quad
-k^2=m_2^2
\quad
{\rm with}
\quad
m_1^2<M^2-\mu^2<m_2^2\,.
\end{align}
Notice here that $m_1^2=\mu^2$ and $m_2^2=\infty$ in the local limit $M^2\to\infty$. The nonlocality therefore induce two effects: First, the nonlocality shifts the mass of $\varphi$ from $\mu^2$ to $m_1^2$, which can be approximated as $\displaystyle m_1^2\simeq\mu^2+\frac{2\mu^4}{M^2}$ in the regime $\mu^2\ll M^2$. The other effect is more drastic. The propagator acquires a new pole at $m_2^2$ near the nonlocal scale. Notably, it has a negative residue (see also Fig.~\ref{fig:tachyon}), which means that the new pole is a ghost mode! We therefore conclude that a ghost mode inevitably appears in this parameter regime, in addition to the ordinary mode with a shifted mass.

\item $\lambda<0$

Finally, let us consider the regime $\lambda<0$ (below we replace $M^2\to-M^2$ in the original action). This parameter regime is not necessarily motivated in the sense that the Euclidean propagator does not have an exponential suppression, but rather has an exponential growth in the UV regime $k^2\gg M^2$. We however discuss the mass spectra for completeness. It is easy to see that the propagator in this regime has a pole at $-k^2=m^2$ ($0<m^2<\mu^2$) with a positive residue. If $\mu^2\ll M^2$, the shifted mass squared is approximated as $\displaystyle m^2\simeq \mu^2-\frac{2\mu^4}{M^2}$.

\end{enumerate}

To summarize, the nonlocality significantly affects the particle spectrum in the tachyon vacuum. In the model~\eqref{nonlocal_tachyon} with the sign choice $M^2>0$ (which is motivated by the mild UV behavior of the propagator), we found two types of spectra: (a) there are no physical particle excitations ($\lambda>\lambda_*$), (b) there appears an undesired ghost mode near the nonlocal scale ($0<\lambda<\lambda_*$). In particular, the latter situation will be problematic if we would like to think of the nonlocal field theory as a UV complete theory, even though there is no problem as long as we regard it, e.g., as a toy model for string field theory or an effective field theory with a cutoff below the mass scale of the ghost. In the rest of this paper, we extend the argument to the Higgs mechanism in nonlocal gauge theories and discuss if it is possible to have a massive gauge boson in a consistent way.

\section{Higgs mechanism in a nonlocal Abelian gauge theory}
\setcounter{equation}{0}
\label{Sec:Abelian}

As an illustrative example for the Higgs mechanism in nonlocal field theories, let us first consider a nonlocal theory of an Abelian gauge field $A_\mu$ and a complex charged scalar $\Phi$:\footnote{
For simplicity, we have set the same nonlocal scale for the gauge field and the scalar, and considered a local $|\Phi|^4$ interaction, although our qualitative argument should hold more generally. Also, since covariant derivatives are non-commutative, the nonlocal action cannot be determined uniquely. For example, we could covariantize the nonlocal kinetic term of the scalar as $\displaystyle-\Phi^\dagger e^{-\frac{D^2}{M^2}}D^2\Phi$ instead. We however concentrate on the model~\eqref{nonlocal_A} as an illustrative example.}
\begin{align}
\label{nonlocal_A}
S=\int d^4x\left[
-\frac{1}{4}F_{\mu\nu}e^{-\frac{\Box}{M^2}}F^{\mu\nu}
-(D_\mu\Phi)^\dagger e^{-\frac{D^2}{M^2}}D^\mu\Phi
+\mu^2\Phi^\dagger e^{-\frac{D^2}{M^2}}\Phi
-\alpha|\Phi|^4
\right]\,,
\end{align}
where the covariant derivative is $D_\mu=\partial_\mu -ig A_\mu$. Similarly to the scalar model of the previous section, we attached the exponential factors so that the propagators of fluctuations around $\Phi=A_\mu=0$ have the same poles as the local theory and the exponential factor gives an exponential suppression of propagators for $k^2\gg M^2$. Also the $\Box$'s in the exponential factor of the charged scalar are gauge covariantized. If the scalar has a tachyonic mass $m_{\rm tachyon}^2=-\mu^
2<0$, this model accommodates a one-parameter family of vacua,
\begin{align}
\left\langle|\Phi|^2\right\rangle=\frac{v^2}{2}
\quad
{\rm with}
\quad
v=\frac{\mu}{\alpha^{1/2}}\,.
\end{align}
Let us choose a vacuum $\displaystyle\left\langle\Phi\right\rangle=\frac{v}{\sqrt{2}}$ without loss of generality and consider fluctuations around it:
\begin{align}
\Phi=\frac{1}{\sqrt{2}}\big[v+\sigma(x)\big]e^{ig\pi(x)}
\,,
\end{align}
where $\pi$ is the Nambu-Goldstone (NG) mode and $\sigma$ is the amplitude mode. If we decompose the gauge field $A^\mu$ into the transverse and longitudinal modes as
\begin{align}
 A^\mu=A_{\bot}^\mu+A_{\parallel}^\mu
\quad
{\rm with}
\quad
\partial_\mu A_{\bot}^\mu=0\,,
\quad
A_{\parallel\mu}=\frac{\partial_\mu\partial_\nu}{\Box}A^\nu\,,
\end{align}
the second order action in fluctuations reads
\begin{align}
S_2=\int d^4x\left[
\mathcal{L}_{\rm trans.}+\mathcal{L}_{\rm long.}+\mathcal{L}_{\sigma}
\right]\,,
\end{align}
where  the second order Lagrangians $\mathcal{L}_{\rm trans.}$, $\mathcal{L}_{\rm long.}$, and $\mathcal{L}_{\sigma}$ of the three sectors are
\begin{align}
\label{Pi_trans}
\mathcal{L}_{\rm trans.}&=
\frac{1}{2}A_{\bot \mu}\left[
e^{-\frac{\Box}{M^2}}\left(\Box-m_A^2\right)+\frac{\mu^2}{M^2}m_A^2
\right]A_{\bot}^\mu\,,
\\
\mathcal{L}_{\rm long.}&=
-\frac{m_A^2}{2}\left(A_{\parallel \mu}-\partial_\mu\pi\right)\left[
e^{-\frac{\Box}{M^2}}\left(1+\frac{\mu^2}{\Box}\right)-\frac{\mu^2}{\Box}
\right]\left(A_{\parallel}^\mu-\partial^\mu\pi\right)\,,
\\
\label{Pi_sigma}
\mathcal{L}_{\sigma}&=
\frac{1}{2}\sigma\left[
e^{-\frac{\Box}{M^2}}\left(\Box+\mu^2\right)-3\mu^2
\right]
\sigma\,.
\end{align}
Here we introduced $m_A^2=g^2v^2$, which is the mass of the gauge boson in the local theory.

\medskip
To discuss physical particle spectrum, let us take the unitary gauge $\pi(x)=0$.\footnote{In Appendix~\ref{Appendix:R_xi}, we provide a concrete form of propagators in the $R_\xi$ type gauge and explicitly check that the NG boson and the gauge-fixing ghost indeed decouple in the unitary gauge.}
The propagators $\Pi_{\mu\nu}^{\rm trans.}$, $\Pi_{\mu\nu}^{\rm long.}$, and $\Pi_\sigma$ of the transverse mode, the longitudinal mode, and the amplitude mode are then given by
\begin{align}
\label{Pi_t}
\Pi_{\mu\nu}^{\rm trans.}(k)&=\left(\eta_{\mu\nu}-\frac{k_\mu k_\nu}{k^2}\right)\left[e^{\frac{k^2}{M^2}}(k^2+m_A^2)-\frac{\mu^2}{M^2}m_A^2\right]^{-1}\,,
\\
\label{Pi_l}
\Pi_{\mu\nu}^{\rm long.}(k)&=\frac{k_\mu k_\nu}{k^2}\frac{1}{m_A^2}
\left[
e^{\frac{k^2}{M^2}}\left(1-\frac{\mu^2}{k^2}\right)+\frac{\mu^2}{k^2}
\right]^{-1}\,,
\\
\label{Pi_s}
\Pi_\sigma(k)&=\left[
e^{\frac{k^2}{M^2}}(k^2-\mu^2)+3\mu^2
\right]^{-1}\,.
\end{align}
Notice that in the limit $M^2\to\infty$, they reproduce the correct unitary gauge propagators of the local Abelian Higgs model.

\medskip
The corresponding on-shell conditions may be stated as
\begin{align}
\label{on-shell_A}
\alpha_{\lambda,s}(x)&=\beta_\lambda(x)=\gamma_\lambda(x)=0\,,
\\
\alpha_{\lambda,s}(x)&=e^{-\lambda x}(x-s)+\lambda s\,,
\\
\beta_\lambda(x)&=e^{-\lambda x}(1+x^{-1})-x^{-1}\,,
\\
\gamma_\lambda(x)&=e^{-\lambda x}(x+1)-3\,,
\end{align}
where we rescaled $x=-k^2/\mu^2$ and introduced $\lambda=\mu^2/M^2$ and $s=m_A^2/\mu^2$. Note that the mass $m$ of the on-shell particle is given by $m^2=x\mu^2$ with $x$ being the solution to Eq.~\eqref{on-shell_A}. We then clarify the pole structure of each propagator.

\paragraph{Amplitude mode}

The propagator~\eqref{Pi_s} of the amplitude mode has essentially the same structure as the tachyon vacuum propagator~\eqref{TV_propagator} in the previous section. For this propagator, the critical value $\lambda_*$ is defined through $\lambda_*^{-1}e^{\lambda_*-1}=3$ and approximately given by $\lambda_*\simeq 0.14$. When $\lambda>\lambda_*$, the propagator has no poles and thus there are no particle excitations. If $0<\lambda<\lambda_*$, there appears a ghost mode near the nonlocal scale, in addition to an ordinary particle excitation with a shifted mass squared. For $\lambda<0$,\footnote{
When we consider $\lambda<0$ in the following, we implicitly assume a positive $\mu^2$ and a negative $M^2$, so that the symmetry is spontaneously broken.} there appears one ordinary particle excitation with a shifted mass squared.

\begin{figure}[t]
\begin{center}
\includegraphics[width=74mm, bb=0 0 360 238]{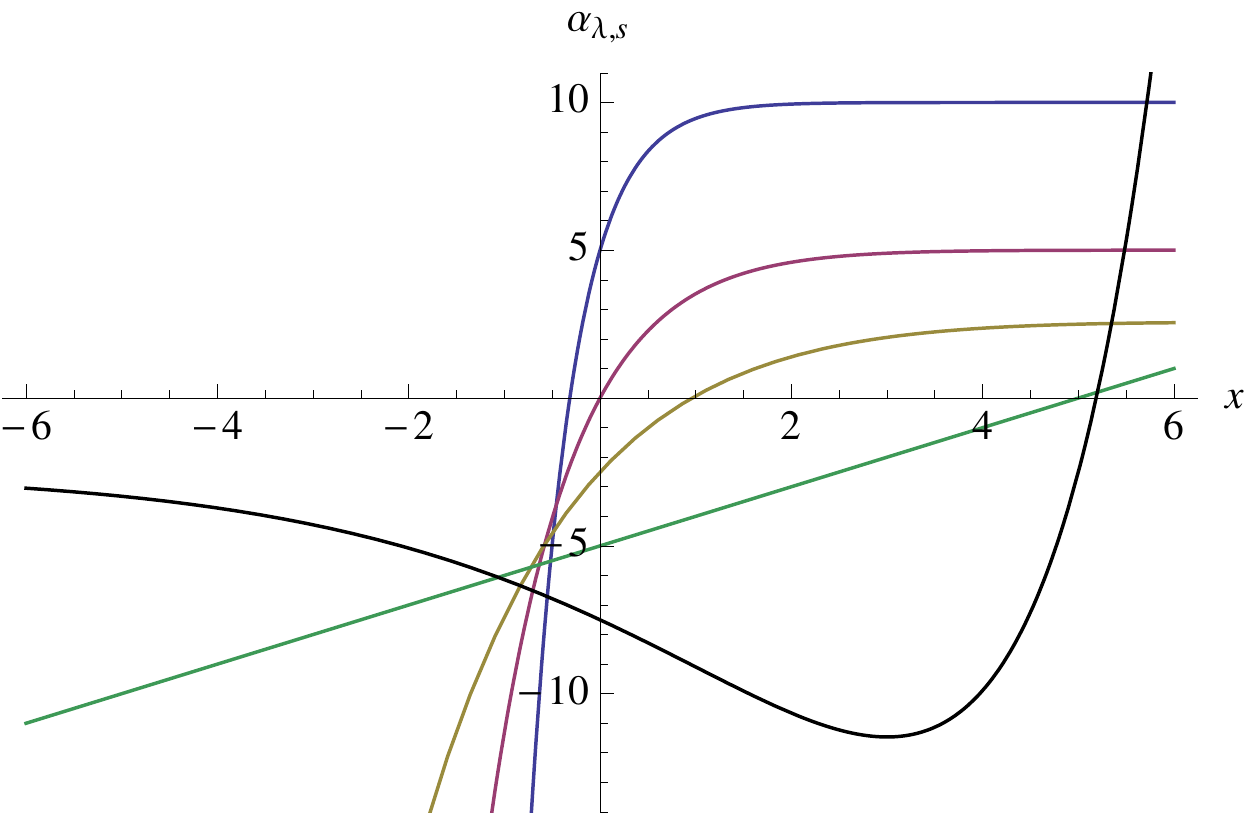}\,\,
\includegraphics[width=74mm, bb=0 0 360 238]{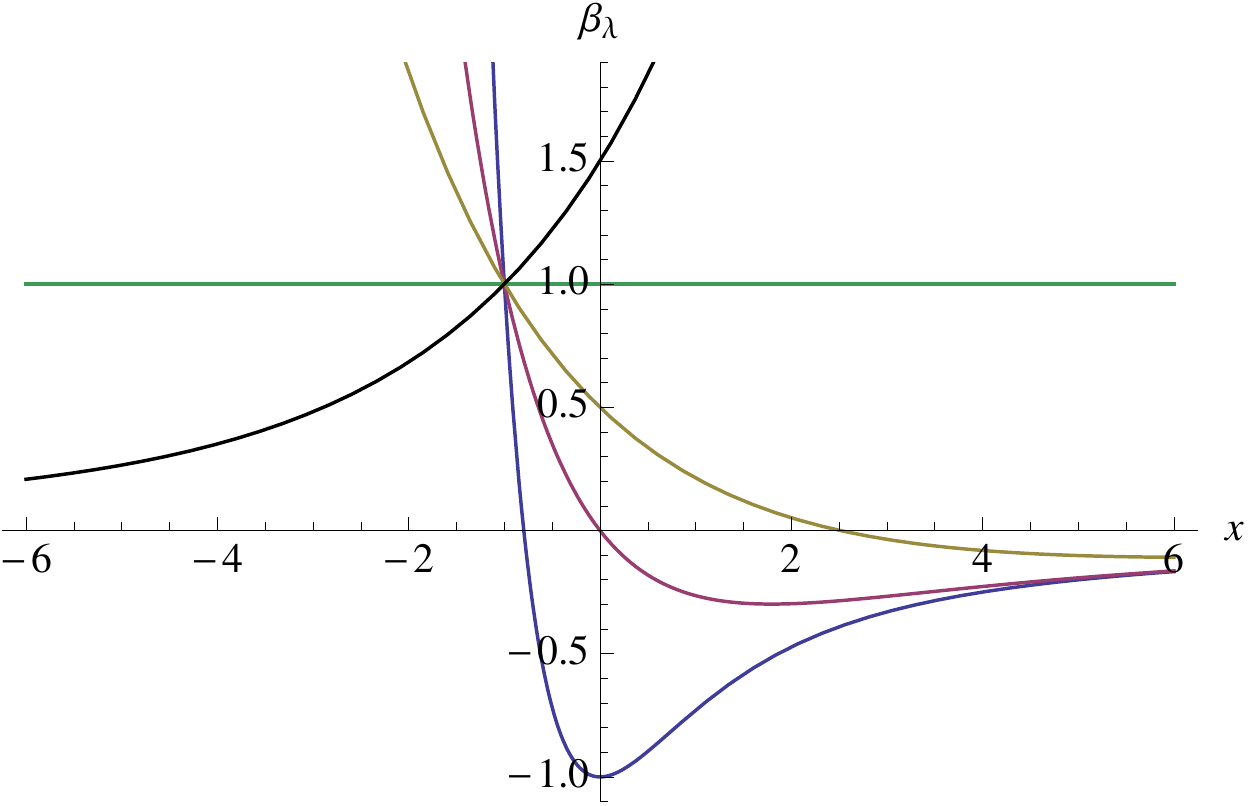}
 \end{center}
\vspace{-5mm}
\caption{
Profile of $\alpha_{\lambda,s}(x)$ and $\beta_{\lambda}(x)$ for $\lambda=2.0$ (blue),
$\lambda=1.0$ (red),
$\lambda=0.5$ (yellow),
$\lambda=0$ (green) 
$\lambda=-0.5$ (black). Here we set $s=5$.}
\label{fig:Abelian}
\end{figure}

\paragraph{Transverse modes}

To understand the pole structure of the propagator~\eqref{Pi_t} of the transverse modes, it is convenient to notice that
\begin{align}
\alpha_{\lambda,s}(0)=s(\lambda-1)\,,
\quad
\left\{\begin{array}{cc}
\displaystyle
\lim_{x\to +\infty}\alpha_{\lambda,s}(x)=\lambda s & \quad{\rm for}\quad\lambda >0\,,
\\[2mm]
\displaystyle
\lim_{x\to -\infty}\alpha_{\lambda,s}(x)=\lambda s
& \quad{\rm for}\quad\lambda <0\,.\end{array}\right.
\end{align}
We then find that there appears one pole with a positive mass squared for $\lambda<1$, whereas there appears one with a tachyonic mass for $\lambda>1$. It is also easy to see that the residue is positive for the former, whereas the latter contains two modes with positive residues and one mode with a negative residue.\footnote{
Notice that eigenvalues of the projector $\displaystyle\eta_{\mu\nu}-\frac{k_\mu k_\nu}{k^2}$ depends on the sign of $-k^2=m^2$: the projector has three positive modes for $-k^2=m^2>0$, whereas it has two positive modes and one negative mode for $-k^2=m^2<0$. This is why the sign of residue is different between $\lambda<1$ and $\lambda>1$, even though the function $\alpha_{\lambda,s}(x)$ has the same sign of gradient at the zero.} See Fig.~\ref{fig:Abelian}. Note that this qualitative property does not depend on $s$, even though the concrete value of the mass depends on $s$.

\paragraph{Longitudinal mode}

Finally, let us look at the propagator~\eqref{Pi_l} of the longitudinal mode. In the local limit $M^2\to\infty$ ($\lambda=0$), $\beta_\lambda(x)$ simply goes to $1$, so that the longitudinal mode has no on-shell excitations as we know. However, the nonlocality significantly affects the physical spectrum.
To understand the pole structure, it is convenient to notice that
\begin{align}
\beta_\lambda(-1)=1\,,\quad
\beta_\lambda(0)=1-\lambda\,,\quad
\left\{\begin{array}{cc}
\displaystyle
\lim_{x\to +\infty}\beta_{\lambda}(x)=-x^{-1}& \quad{\rm for}\quad\lambda >0\,,
\\[2mm]
\displaystyle
\lim_{x\to -\infty}\beta_{\lambda}(x)=-x^{-1}
& \quad{\rm for}\quad\lambda <0\,.\end{array}\right.
\end{align}
We then find that there appears one pole with a positive mass squared for $0<\lambda<1$, one pole with a tachyonic mass for $\lambda>1$,  and no poles for $\lambda<0$. Also, the residue of the pole for $0<\lambda<1$ is negative, whereas that for $\lambda>1$ is positive. See Fig.~\ref{fig:Abelian}

\begin{figure}[t]
\begin{center}
\includegraphics[width=150mm]{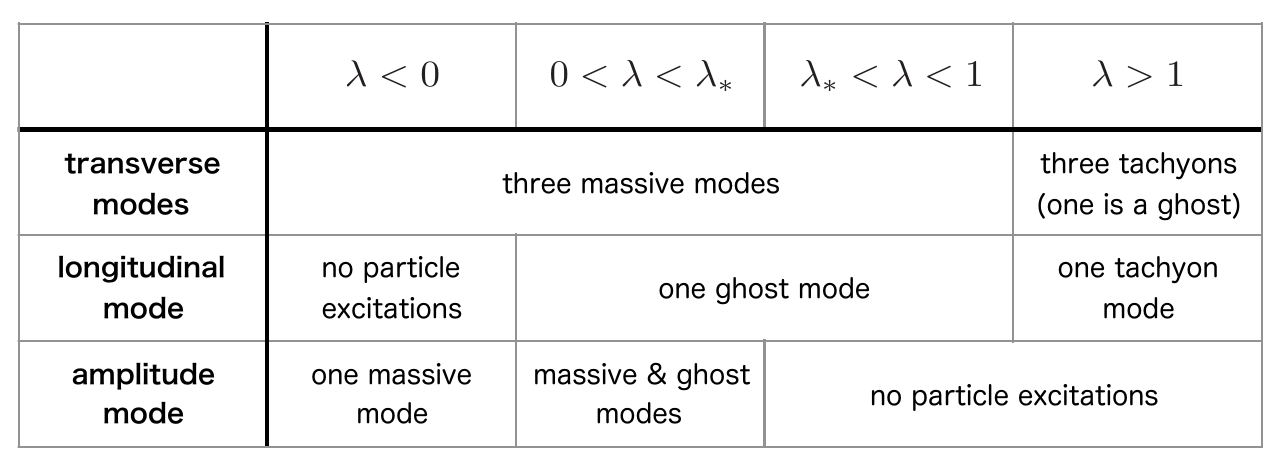} \end{center}
\vspace{-5mm}
\caption{
Summary of the particle spectrum.}
\label{fig:table}
\end{figure}

\bigskip
To summarize, the particle spectrum in the symmetry broken phase of the model~\eqref{nonlocal_A} highly depends on the ratio $\lambda=\mu^2/M^2$ of the mass parameter $\mu$ and the nonlocal scale $M$. As shown in Fig.~\ref{fig:table}, a ghost mode inevitably appears in the simple nonlocal model~\eqref{nonlocal_A} as long as we consider a positive $\lambda$, which is motivated by the mild UV behavior of the propagator. It will be a crucial obstruction to making the simple nonlocal model a UV complete theory. Does it imply that the nonlocal theory cannot accommodate a massive gauge boson in a consistent way? One way out would be to accept $\lambda < 0$ as a possible solution (even though it requires a positive $\mu^2$ and a negative $M^2$) but still one needs to define the contour of the path integral in an unusual way (not an analytic continuation of the Euclidean theory).  Another possible way out would be to impose some symmetry. To explore the latter possibility, in the next section, we discuss a nonlocal Yang-Mills with an adjoint Higgs which is obtained after the KK reduction of the nonlocal pure Yang-Mills in higher dimension. We there demonstrate that it is indeed possible to evade the undesired ghost even for a positive $\lambda$ thanks to the gauge symmetry of the higher dimension.

\section{Nonlocal Yang-Mills with an adjoint Higgs}
\setcounter{equation}{0}
\label{Sec:non-Abelian}

As a simple and illustrative example, we consider a dimensional reduction of a nonlocal extension of the $SU(2)$ Yang-Mills from $6$D to $4$D. As a $6$D nonlocal Yang-Mills, let us consider the following action:\footnote{
Similarly to the Abelian case, there is an ambiguity when covariantizing the derivatives. Even though we focus on the model~\eqref{6D_YM} for illustration, it is straightforward to extend the argument to a general setup and show that the particle spectrum does not depend on the details of the covariantization scheme.}
\begin{equation}
\label{6D_YM}
S=\int d^{6}x\,\mathrm{Tr}\left[
-\frac{1}{4}F_{MN}\,e^{-\frac{D^2}{M^2}}F^{MN}
\right]\,,
\end{equation}
where $M$ is the non-local scale. The covariant derivative is $D_M\,*\,=\partial_M\,*\,-ig[A_M,\,*\,]$ and its square $D^2=D_MD^M$ is
\begin{equation}
\label{eq:10dim_squared_covariant_derivative}
D^2\,\ast=\Box\,*\,-ig\left[\partial_MA^M,\,\ast\,\right]-2ig\left[A^M,\partial_M\,\ast\,\right]-g^2\left[A_M,\left[A^M,\,\ast\,\right]\right]\,,
\end{equation}
where $\ast$ stands for adjoint fields.
After dimensional reduction, we obtain a non-local 4D Yang-Mills with two adjoint scalars:
\begin{align}
\label{4DYM}
S=\int d^4x\,\mathrm{Tr}\left[-\frac{1}{4}F_{\mu\nu}\,e^{-\tfrac{\mathcal{D}^2}{M^2}}F^{\mu\nu}
-\frac{1}{2}D_\mu\phi_I\,e^{-\frac{\mathcal{D}^2}{M^2}}D^\mu\phi_I
+\frac{g^2}{4}[\phi_I,\phi_J]\,e^{-\frac{\mathcal{D}^2}{M^2}}[\phi_I,\phi_J]\right]\,,
\end{align}
where $I,J=4,5$. We emphasize that the exponent is $\mathcal{D}^2$ defined by
\begin{align}
\mathcal{D}^2\,\ast=\Box\,\ast-ig\left[\partial_\mu A^\mu,\,\ast\,\right]-2ig\left[A^\mu,\partial_\mu\,\ast\,\right]-g^2\left[A_\mu,\left[A^\mu,\,\ast\,\right]\right]
-g^2\left[\phi_I,\left[\phi_I,\,\ast\,\right]\right]\,,
\end{align}
rather than the $4$D covariant derivative squared,
\begin{align}
D_\mu D^\mu\,\ast=\Box\,\ast-ig\left[\partial_\mu A^\mu,\,\ast\,\right]-2ig\left[A^\mu,\partial_\mu\,\ast\,\right]-g^2\left[A_\mu,\left[A^\mu,\,\ast\,\right]\right]\,.
\end{align}

\medskip
Let us then consider fluctuations around a vacuum,
\begin{align}
\langle \phi_4\rangle=0\,,
\quad
\langle\phi_5\rangle=vT_3\,,
\quad
\varphi_I=\phi_I-\langle\phi_I\rangle
\quad
(I=4,5)\,,
\end{align}
where $T_a$ $(a=1,2,3)$ denotes the $SU(2)$ generators and we use the convention,
\begin{align}
[T_a,T_b]=i\epsilon_{abc}T_c\,,
\quad
{\rm Tr}(T_a\,T_b)=\delta_{ab}\,.
\end{align}
This vacuum spontaneously breaks the $SU(2)$ gauge symmetry to $U(1)$. The second order action in the fluctuation is
\begin{align}
\nonumber
S_2&=\int d^4x\,\mathrm{Tr}\bigg[-\frac{1}{4}F_{\mu\nu}\,e^{-\tfrac{\overline{\mathcal{D}}^2}{M^2}}F^{\mu\nu}
+\frac{g^2}{2}[\langle\phi_5\rangle,A^\mu]\,e^{-\frac{\overline{\mathcal{D}}^2}{M^2}}[\langle\phi_5\rangle,A^\mu]
-ig\partial_\mu\varphi_5\,e^{-\frac{\overline{\mathcal{D}}^2}{M^2}}[\langle\phi_5\rangle,A^\mu]
\\
\label{before_gauge_fixing}
&\qquad\quad
-\frac{1}{2}\partial_\mu\varphi_4\,e^{-\frac{\overline{\mathcal{D}}^2}{M^2}}\partial^\mu\varphi_4
+\frac{g^2}{2}[\langle\phi_5\rangle,\varphi_4]\,e^{-\frac{\overline{\mathcal{D}}^2}{M^2}}[\langle\phi_5\rangle,\varphi_4]
-\frac{1}{2}\partial_\mu\varphi_5\,e^{-\frac{\overline{\mathcal{D}}^2}{M^2}}\partial^\mu\varphi_5
\bigg]\,,
\end{align}
where we introduced
\begin{align}
\overline{\mathcal{D}}^2\,\ast=\Box
-g^2\left[\langle\phi_5\rangle,\left[\langle\phi_5\rangle,\,\ast\,\right]\right]\,.
\end{align}
Note that the quadratic terms in $F_{\mu\nu}$ do not contribute to the second order action even though we used $F_{\mu\nu}$ in the above for notational simplicity.

\paragraph{Nonlocal extension of the $R_\xi$ gauge:}

It is very useful to introduce a gauge fixing condition that eliminates the mixing of $\varphi_5$ and $A_\mu$ in the action~\eqref{before_gauge_fixing}. This is implemented with a gauge-fixing function,
\begin{equation}
	\label{eq:non-local_gauge_fixing_function}
	G=\frac{1}{\sqrt{\xi}}\left(e^{-\frac{\overline{\mathcal{D}}^2}{2M^2}}\partial_\mu A^\mu-ig\xi\,e^{-\frac{\overline{\mathcal{D}}^2}{2M^2}}[\langle\phi_5\rangle,\varphi_5]\right)\,,
\end{equation}
which reduces to the $R_\xi$ gauge in the local limit $M^2\to\infty$. The gauge fixing term modifies the second order action as
\begin{align}
\nonumber
\widetilde{S}_2&=\int d^4x\,\mathrm{Tr}\bigg[-\frac{1}{4}F_{\mu\nu}\,e^{-\tfrac{\overline{\mathcal{D}}^2}{M^2}}F^{\mu\nu}
-\frac{1}{2\xi}\partial_\mu A^\mu e^{-\frac{\overline{\mathcal{D}}^2}{M^2}}\partial_\mu A^\mu
+\frac{g^2}{2}[\langle\phi_5\rangle,A^\mu]\,e^{-\frac{\overline{\mathcal{D}}^2}{M^2}}[\langle\phi_5\rangle,A^\mu]
\\*
\nonumber
&\qquad\qquad\qquad\!
-\frac{1}{2}\partial_\mu\varphi_4\,e^{-\frac{\overline{\mathcal{D}}^2}{M^2}}\partial^\mu\varphi_4
+\frac{g^2}{2}[\langle\phi_5\rangle,\varphi_4]\,e^{-\frac{\overline{\mathcal{D}}^2}{M^2}}[\langle\phi_5\rangle,\varphi_4]
\\*
&\qquad\qquad\qquad\!
-\frac{1}{2}\partial_\mu\varphi_5\,e^{-\frac{\overline{\mathcal{D}}^2}{M^2}}\partial^\mu\varphi_5
+\frac{\xi g^2}{2}[\langle\phi_5\rangle,\varphi_5]\,e^{-\frac{\overline{\mathcal{D}}^2}{M^2}}[\langle\phi_5\rangle,\varphi_5]
\bigg]\,,
\end{align}
where  $\displaystyle\widetilde{S}_2=S_2+\int d^4x\,\mathrm{Tr}[-\tfrac{1}{2}G^2]$. The corresponding quadratic action of the gauge-fixing ghost $c$ is
\begin{align}
S_{2,{\rm ghost}}=\int d^4x\,
\mathrm{Tr}\left[\bar{c}e^{-\frac{\overline{\mathcal{D}}^2}{2M^2}}\Box c-\xi g^2\bar{c}\,e^{-\frac{\overline{\mathcal{D}}^2}{2M^2}}\left[\langle\phi_5\rangle,\left[\langle\phi_5\rangle,c\right]\right]\right]\,,
\end{align}
where we omitted cubic and higher order interaction terms between the gauge-fixing ghost and other fields because they are not relevant to the particle spectrum.

\paragraph{Propagators:}
Finally, let us derive the propagators. First, the $a=3$ sector is not affected by the condensate, so that the propagators are
\begin{align}
\label{propagator_NA1}
\Pi^{(3)}_{\mu\nu}(k)=\frac{e^{-\frac{k^2}{M^2}}}{k^2}\left[\eta_{\mu\nu}-(1-\xi)\frac{k_\mu k_\nu}{k^2}\right]\,,
\quad
\Pi^{(3)}_{I}(k)=\Pi^{(3)}_c(k)=\frac{e^{-\frac{k^2}{M^2}}}{k^2}\,,
\end{align}
where $\Pi^{(a)}_{\mu\nu}$, $\Pi^{(a)}_{I}$, and $\Pi^{(a)}_{c}$ are the propagators of gauge fields, $\varphi_I$'s, and the ghosts, respectively. The superscript $(a)$ indicates that the propagator has an $SU(2)$ index $a$. To obtain the propagators in the $a=1,2$ sector, it is convenient to note the relation,
\begin{align}
\left[\langle\phi_5\rangle,\left[\langle\phi_5\rangle,T_a\right]\right]=v^2T_a &\quad \text{for}\quad a=1,2\,.
\end{align}
It is then easy to find
\begin{align}
\label{propagator_NA2}
&\Pi^{(1,2)}_{\mu\nu}(k^2)=\frac{e^{-\frac{k^2+m^2}{M^2}}}{k^2+m^2}\biggl(\eta_{\mu\nu}-(1-\xi)\frac{k_\mu k_\nu}{k^2+\xi m^2}\biggr)\,,\\
\label{propagator_NA3}
&\Pi^{(1,2)}_{4}(k^2)=\frac{e^{-\frac{k^2+m^2}{M^2}}}{k^2+m^2}\,,
\quad
\Pi^{(1,2)}_5(k^2)=\frac{e^{-\frac{k^2+m^2}{M^2}}}{k^2+\xi m^2}\,,
\quad
\Pi^{(1,2)}_c(k^2)=\frac{e^{-\frac{k^2+m^2}{2M^2}}}{k^2+\xi m^2}\,,
\end{align}
where we introduced $m^2=g^2v^2$. The point is that the exponential factors in the propagators~\eqref{propagator_NA1},~\eqref{propagator_NA2}, and~\eqref{propagator_NA3} are factorized, so that the pole structure is the same as the local case, even though the exponential suppression of the propagators is still there.\footnote{
One might wonder if a similar factorization may occur even in the scalar example discussed in Sec.~\ref{Sec:tachyon} by modifying the $\phi^3$ interaction as $(e^{-\Box/2M^2}\phi)^3$. However, this theory is reduced to a local one by a field redefinition $e^{-\Box/2M^2}\phi\to\phi$. A similar argument holds also for the scalar QED example discussed in Sec.~\ref{Sec:Abelian}.
On the other hand, it is not difficult to check that the non-Abelian setup~\eqref{4DYM} cannot be reduced to a local theory by a field redefinition. Indeed, we can explicitly show that on-shell scattering amplitudes are different from the local ones and thus the theory is nonlocal.}

\medskip
We here emphasize that the factorization of the exponential factors is a consequence of the $6$D gauge symmetry: As long as we consider the $4$D Yang-Mills with adjoint scalars, there is no reason to expect that the same exponential factor appears in all the three terms of Eq.~\eqref{4DYM}. For example, the quartic interactions of the scalar could be local, just as in the models discussed in the previous sections. Also, the nonlocal scale $M$ in the exponent could be not exactly the same among the three terms. In this way, the $6$D gauge symmetry naturally realizes the factorization, which evades the undesired ghost mode in the symmetry broken phase, without spoiling the mild UV behavior of propagators.

\section{Conclusion}
\setcounter{equation}{0}
\label{Sec:conclusion}

In this paper we studied spontaneous gauge symmetry breaking and the Higgs mechanism in a class of nonlocal gauge theories. We found that the nonlocality drastically changes the particle spectrum in the symmetry broken phase, similarly to the open string tachyon condensation case. In particular, there may potentially appear an undesirable ghost near the nonlocal scale. It will be a crucial obstruction to making the simple nonlocal model a UV complete theory, hence absence of such ghosts can be used as a consistency requirement of the theory. We then demonstrated that it is indeed possible to evade the undesired ghost by imposing an extra symmetry such as gauge symmetries in higher dimensional spacetime.

\medskip
There are several future directions worth exploring. First, it is important to extend our study to the full open string field theory. Even though it seems technically complicated, we believe that recent progress in the numerical~\cite{Matej,Karczmarek:2012pn,Kudrna:2012um,Kudrna:2016ack} and analytic~\cite{Schnabl:2005gv,Ellwood:2006ba,Kiermaier:2007ba,Schnabl:2007az,Fuchs:2007yy,Kiermaier:2007vu,Erler:2007rh,Okawa:2007ri,Fuchs:2007gw,Kiermaier:2007ki,Kiermaier:2010cf,Bonora:2010hi,Erler:2011tc,Noumi:2011kn,Takahashi:2011wk,Hata:2011ke,Murata:2011ep,Masuda:2012kt,Erler:2014eqa,Ishibashi:2016xak,Kojita:2016jwe,Ishibashi:2018ynb} construction of classical solutions will be useful. For example, numerical solutions for bosonic open string field theory on intersecting D-branes were studied in the level truncation approach~\cite{Matej}. It would be very nice if we could generalize it to the superstring and discuss the Standard Model Higgs mechanism in a realistic D-brane setup. It will also be interesting to clarify how string field theory evades the undesired ghost we encountered in the simple nonlocal model.
In string field theory, there exist infinitely many massive higher spin fields in addition to light fields. Furthermore, they enjoy an infinite dimensional gauge symmetry associated with the conformal symmetry of the worldsheet theory~\cite{Gaberdiel:1997ia,Zwiebach:1992ie}. We expect that this large gauge symmetry plays a crucial role in making the theory healthy.
A more challenging but very important direction will be the construction of an analytic solution for D-brane recombination. We believe that the general construction in Ref.~\cite{Erler:2014eqa} will shed  new light on this direction. Second, it would be interesting to carry out further studies of the consistency requirements of nonlocal field theories. The subtlety associated with the ghost we found in this paper appeared even at the tree-level. It will be important to investigate a similar issue at the loop-level. We hope to report our progress in these directions elsewhere.

\section*{Acknowledgements}
HI is supported in part by the ``CUniverse'' research promotion project by Chulalongkorn University (grant reference CUAASC).
TN  is in part supported by JSPS KAKENHI Grant Numbers JP17H02894 and JP18K13539, and MEXT KAKENHI Grant Number JP18H04352.
GS is supported in part by the DOE grant DE-SC0017647 and the Kellett Award of the University of Wisconsin.
PS is supported by the DFG Transregional Collaborative Research Centre TRR 33 ``The Dark Universe''.

\appendix

\section{$R_\xi$ gauge propagators in nonlocal Abelian gauge theory}
\setcounter{equation}{0}
\label{Appendix:R_xi}

In analogy with the $R_\xi$ gauge, let us introduce two types of one-parameter families of gauge-fixing functions:
\begin{align}
\label{G1_gauge}
G&=
\frac{1}{\sqrt{\xi}}\Bigg[\left[
e^{-\frac{\Box}{M^2}}\left(1+\frac{\mu^2}{\Box}\right)-\frac{\mu^2}{\Box}\right]
\partial_\mu A^\mu -\xi m_A^2\pi\Bigg]\,,
\\
\label{G2_gauge}
\widetilde{G}&=
\frac{1}{\sqrt{\xi}}\Bigg[\partial_\mu A^\mu
-\xi m_A^2\left[
e^{-\frac{\Box}{M^2}}\left(1+\frac{\mu^2}{\Box}\right)-\frac{\mu^2}{\Box}\right]\pi\Bigg]
\,,
\end{align}
both of which reproduce the ordinary $R_\xi$ gauge in the local limit $M\to\infty$. The first one reproduces the unitary gauge in the limit $\xi\to\infty$, whereas the second one gives the Lorentz gauge in the limit $\xi=0$.
Below, we derive an explicit form of propagators in each gauge. Note that we do not discuss the propagators of the transverse modes and the amplitude mode in this Appendix because they are gauge-invariant and therefore given by Eq.~\eqref{Pi_trans} and Eq.~\eqref{Pi_sigma} in any gauge.

\paragraph{The $G$ gauge propagators}
Let us begin with the gauge~\eqref{G1_gauge}. First, the gauge-fixing term modifies the Lagrangian of the longitudinal sector as
\begin{align}
\mathcal{L}_{\rm long.}-\frac{G^2}{2}
&=\frac{1}{2\xi}A_{\parallel \mu}\left[
e^{-\frac{\Box}{M^2}}\left(1+\frac{\mu^2}{\Box}\right)-\frac{\mu^2}{\Box}
\right]
\left[
e^{-\frac{\Box}{M^2}}\left(\Box+\mu^2\right)-\mu^2-\xi m_A^2
\right]
A_{\parallel}^\mu
\nonumber
\\
&\quad
+
\frac{m_A^2}{2}\pi\left[
e^{-\frac{\Box}{M^2}}\left(\Box+\mu^2\right)-\mu^2-\xi m_A^2
\right]\pi\,.
\end{align}
The propagator of the NG bosons is then
\begin{align}
\Pi_\pi(k)=\frac{1}{m_A^2}\Pi_\xi(k)
\quad
{\rm with}
\quad
\Pi_\xi(k)=\bigg[{e^{\frac{k^2}{M^2}}(k^2-\mu^2)+\mu^2+\xi m_A^2}\bigg]^{-1}\,,
\end{align}
where the prefactor $\frac{1}{m_A^2}$ is due to our normalization. The momentum dependent part $\Pi_\xi(k)$ reproduces a scalar propagator with a gauge dependent mass $\xi m_A^2$ in the local limit $k^2\ll M^2$, but its pole structure is modified by the nonlocality in general. In the limit $\xi\to\infty$, the pole of the propagator disappears and also the propagator vanishes, so that the NG boson decouples. This is the unitary gauge we discussed in the main text.

\medskip
On the other hand, the propagator of the longitudinal mode $A_{\parallel\mu}$ is given by
\begin{align}
\label{P_long_xi}
\Pi_{\mu\nu}^{\rm long.}(k)=\frac{k_\mu k_\nu}{k^2}f(k^2)\,\xi\,\Pi_\xi(k)
\quad
{\rm with}
\quad
f(k^2)={k^2}\bigg[{e^{\frac{k^2}{M^2}}(k^2-\mu^2)+\mu^2}\bigg]^{-1}\,,
\end{align}
which has two types of poles: The first is those of the function $\Pi_\xi$, which have the same gauge-dependent  masses as the NG boson $\pi$. In particular, they disappear in the unitary gauge $\xi\to\infty$. The second type is those arising from the prefactor $f(k^2)$. It is simply $1$ in the local limit $k^2\ll M^2$, but the nonlocality brings about additional poles. The poles we discussed in the main text are these second type of poles.

\medskip
Finally, we derive the propagator of the gauge-fixing ghost. First, the gauge transformation of the gauge-fixing function~\eqref{G1_gauge} reads
\begin{align}
\delta G=\frac{1}{\sqrt{\xi}}\left[
e^{-\frac{\Box}{M^2}}\left(\Box+\mu^2\right)-\mu^2 -\xi m_A^2\right]\alpha
\,,
\end{align}
where we used
\begin{align}
A_\mu\to A_\mu+\partial_\mu\alpha\,,
\quad
\pi\to\pi+\alpha\,.
\end{align}
The corresponding ghost Lagrangian is therefore
\begin{align}
\mathcal{L}_{\rm ghost}&=\bar{c}
\left[
e^{-\frac{\Box}{M^2}}\left(\Box+\mu^2\right)-\mu^2
-\xi m_A^2
\right]
c
\,.
\end{align}
The propagator of the ghost is then
\begin{align}
\Pi_c(k)=\Pi_\xi(k)\,,
\end{align}
which is nothing but the NG boson propagator up to a numerical normalization factor. In particular, the ghost decouples in the unitary gauge $\xi\to\infty$ just as the NG boson does.

\paragraph{The $\widetilde{G}$ gauge propagators}

We then move on to the second gauge~\eqref{G2_gauge}.
In this gauge, the Lagrangian of the longitudinal sector is modified as
\begin{align}
\mathcal{L}_{\rm long.}-\frac{\widetilde{G}^2}{2}
&=\frac{1}{2\xi}A_{\parallel \mu}\Big[
\Box-\xi m_A^2\widetilde{f}(-\Box)
\Big]
A_{\parallel}^\mu
\nonumber
\\
&\quad
+
\frac{m_A^2}{2}\pi
\widetilde{f}(-\Box)
\Big[
\Box-\xi m_A^2\,\widetilde{f}(-\Box)
\Big]\pi\,,
\end{align}
where the function $\widetilde{f}$ is defined by $\widetilde{f}=1/f$.
The propagators of the NG boson and the longitudinal mode of the gauge boson are
\begin{align}
\widetilde{\Pi}_\pi(k)=\frac{1}{m_A^2}{f(k^2)}\widetilde{\Pi}_\xi(k)\,,
\quad
\widetilde{\Pi}_{\mu\nu}^{\rm long.}=\frac{k_\mu k_\nu}{k^2}\xi \,\widetilde{\Pi}_\xi(k)\,,
\end{align}
where we introduced
\begin{align}
\widetilde{\Pi}_\xi(k)=\frac{1}{k^2+\xi m_A^2\widetilde{f}(k^2)}\,.
\end{align}
We then derive the propagator of the gauge-fixing ghost. First, the gauge transformation of the gauge-fixing function~\eqref{G2_gauge} reads
\begin{align}
\delta \widetilde{G}=\frac{1}{\sqrt{\xi}}\left[
\Box -\xi m_A^2\widetilde{f}(-\Box)\right]\alpha
\,,
\end{align}
so that the corresponding ghost Lagrangian is
\begin{align}
\mathcal{L}_{\rm ghost}&=\bar{c}
\left[
\Box
-\xi m_A^2\widetilde{f}(-\Box)
\right]
c
\,.
\end{align}
The propagator of the ghost is then
\begin{align}
\widetilde{\Pi}_c(k)=\widetilde{\Pi}_\xi(k)\,.
\end{align}
As we mentioned earlier, the Lorentz gauge corresponds to the limit $\xi=0$. In the Lorentz gauge $\xi=0$, we find that $\widetilde{\Pi}_\xi(k)$ and therefore the ghost propagator $\widetilde{\Pi}_c(k)$ reduce to a massless propagator. The NG boson propagator $\widetilde{\Pi}_\pi$ is a massless propagator with a momentum-dependent factor $f(k^2)$, which gives rise to extra poles in addition to the massless pole. Also the propagator $\widetilde{\Pi}_{\mu\nu}^{\rm long.}$ of the longitudinal mode vanishes.

\bibliography{NLH}{}
\bibliographystyle{utphys}

\end{document}